\documentclass{article}
\usepackage{appendix}
\usepackage{amsmath}
\usepackage[retainorgcmds]{IEEEtrantools}
\usepackage{graphicx}
\usepackage{booktabs}
\usepackage{float}
\restylefloat{table}

\begin{document}
\emph{The following article has been accepted by \emph{The American Journal of Physics}. After it is published, it will be found at Link:} \textbf{https://aapt.scitation.org/journal/ajp}
\title{Why pushing a bell does not produce a sound}
\author{Loris Ferrari \\ Department of Physics and Astronomy (DIFA) of the University \\via Irnerio, 46 - 40126, Bologna,Italy}
\maketitle
\begin{abstract}
The difference between \textquoteleft beating\textquoteright$\:$and \textquoteleft pushing\textquoteright$\:$ results in the perception that a push just makes the object move as a whole, while a beat produces also a sound. Through a detailed analysis of the physics underlying such everyday experiences, we identify the \emph{strength}, the \emph{duration} and the \emph{softness} of the applied contact force, as the main (measurable) characteristics that mark such difference. The strength determines the final velocity $\Delta v$ achieved by the body. The duration $2\tau$ compares to the time $\tau_\ell$ the sound takes to cross the body. The softness $\gamma$ (a positive exponent) results from the shape in time of the contact force. Those three elements enter the formula for the intensity of the sound produced. The relevant role of the softness is stressed and specific values are calculated for a thin metallic bar, chosen as the simplest possible model system. \newline       
\end{abstract}
e-mail: loris.ferrari@unibo.it
telephone: ++39-051-2091136

\section{Introduction}
\label{intro}
The title of the present work aims to stimulate some curiosity, about phenomena whose physical complexity may be hidden by the screen of everyday habits. Beating a metallic object (skillet, bar or bell) with a hammer, produces a sound. Pushing or pulling the same object, instead, generates just a movement. Those everyday experiences are so obvious, that asking \textquoteleft why?\textquoteright$\:$might look bizarre, if not nonsense. Physicists, however, know that the motion equations in the two cases are exactly the same, apart from the applied forces. So, a spontaneous question arises, about what, in the applied forces, does mark the difference. One aspect is certainly the \emph{duration} of the contact between the object and the external agents: a \textquoteleft short\textquoteright$\:$contact (beat) mainly result in a sound, while a \textquoteleft long\textquoteright$\:$contact (push) does make the system move \emph{as a whole}, without sound production, but why? In an attempt to give an answer, the physicist is now triggered on a further important aspect: since \textquoteleft short\textquoteright$\:$or \textquoteleft long\textquoteright$\:$does not mean anything, without a characteristic time scale to compare with, one expects that this time scale should emerge naturally from the motion equations. Actually, the sound is a pressure wave in the surrounding air, generated by the dissipation of the internal energy of the object, which is elastic, in a first approximation. A consequence of elasticity is the existence of a longitudinal and transversal sound velocity, $c_{l}$ and $c_{t}$, which result in two characteristic times $\tau_{l}$ and $\tau_t$ (comparable in magnitude) for the sound to cross the object. At this stage, it is not difficult to guess that $\tau_{l}$ and/or $\tau_t$ are the time scales to which the duration of the contact force is to be compared.

Another experimental evidence is that a hammer beat produces a much louder sound than a soft mallet drumstick, despite the intensity and the duration of the contacts are comparable. So, the \emph{softness} of the contact is relevant too, once the term \textquoteleft softness\textquoteright$\:$ has been given a precise physical meaning.  

In an attempt to give a quantitative picture of the \textquoteleft body perceptions\textquoteright$\:$ described above, one has to study the dynamics of an elastic, \emph{unconstrained} object, initially at rest, experiencing a contact force of variable duration, on a small portion of its surface. The amplitude of the free internal oscillations, after the end of the contact, will determine the elastic energy to be transformed into sound and the way it depends on the characteristics of the force itself. The present work aims to approach the problem in the simplest possible case of an elastic thin bar (Section \ref{Bar}) of small transversal section (Fig.1). 

\begin{figure}[htbp]
\centering
\includegraphics[width=6in]{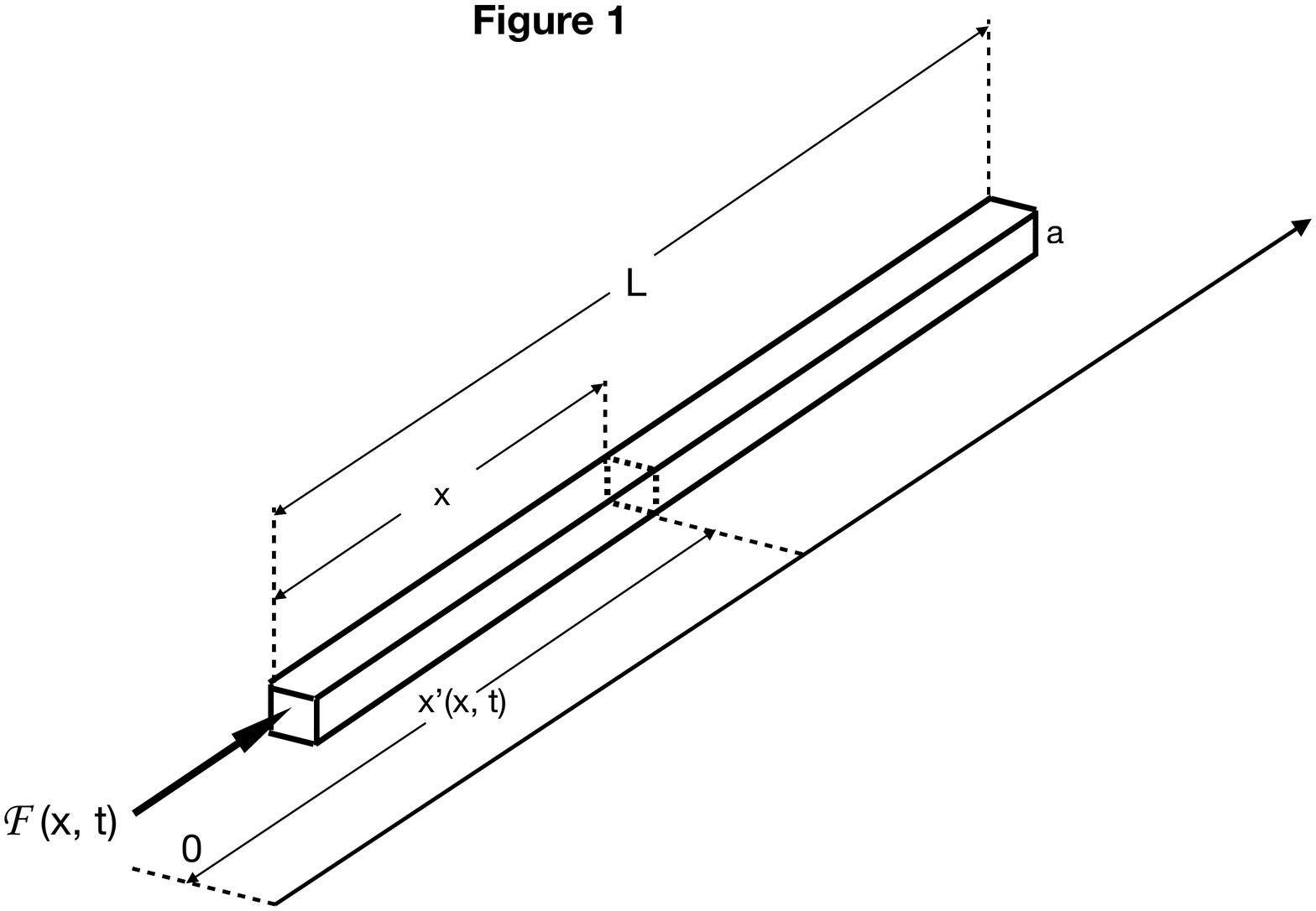}
\caption{\textbf{Schematic of the thin bar}. Notice that the actual position $x'(x,\:t)$ of the transversal section includes the displacement from the origin of the laboratory frame of reference.}
\label{default}
\end{figure} 

As we shall see, the bar's problem reduces to the problem of a harmonic oscillator, under the action of a transient driving (Section \ref{1osc}). For a time-dependent driving, expandible in a generalized non-negative power series (Fig.2), 

\begin{figure}[htbp]
\centering
\includegraphics[width=4in]{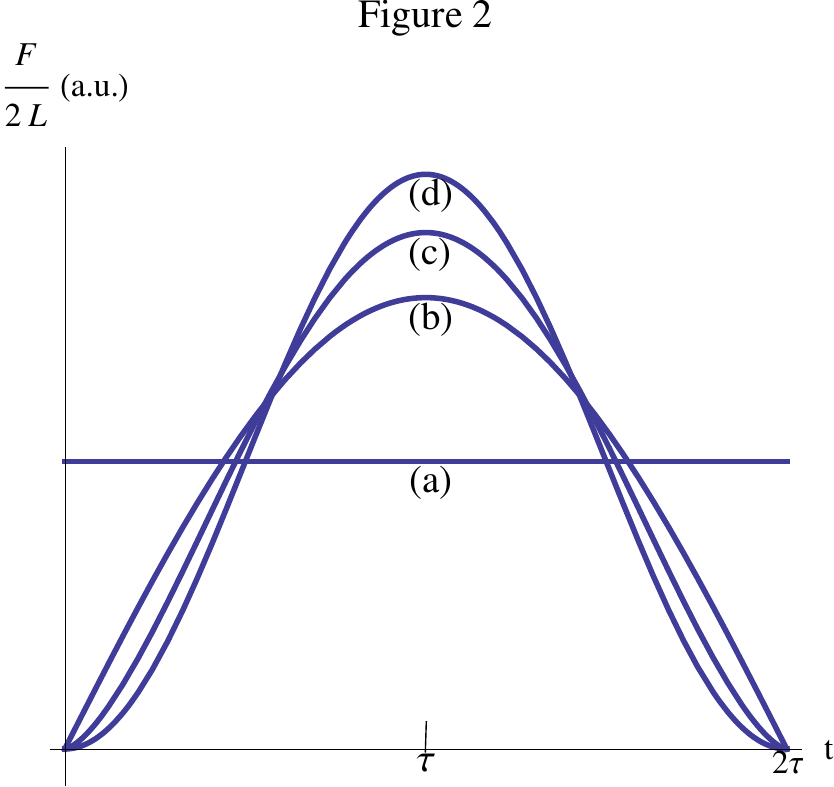}
\caption{\textbf{Transient contact force}. The forces are proportional to $[\sin(\pi t/(2\tau))]^{\alpha_0}$ and normalized to the same value, according to eq.~\eqref{intPsi}. The parameter $\alpha_0=0$ (a), $1$ (b), $\frac{3}{2}$ (c), $2$ (d) is the smallest exponent in the series expansion eq.~\eqref{Psi}. Cases (b) and (c) correspond to ref.s \cite{Hunter, Reed} .}
\label{default}
\end{figure}

the solutions (Sections \ref{Bar2}, \ref{tau->inf}) will illustrate all the possible cases, between the (almost) pure production of a sound and the (almost) pure movement of the centre of mass (c.m.). In Section \ref{soundint}, it is shown that the three main elements of the contact (strength, duration, softness) yield an upper limiting value $\mathcal{I}_M$ for the sound intensity:

\begin{equation}
\label{Ab}
\mathcal{I}_{snd}(t)<\mathcal{I}_M\propto(\Delta v)^2\left(\frac{\tau_l}{\tau}\right)^\gamma\:,
\end{equation}
\\
where $\Delta v$ is the velocity achieved by the c.m., $2\tau$ is the duration of the contact, and $\tau_l$ is the time the (longitudinal) elastic wave takes to cross the bar. Quantitative estimates of $\mathcal{I}_M$ (Table 2) will be obtained for different shapes of the contact force (Fig. 2), acting on a steel bar (Table 1).

Equation \eqref{Ab} supports the physical intuition that the larger the ratio between the duration $2\tau$, and the crossing time $\tau_l$, the stronger the attenuation of the resulting sound, determined by the positive integer exponent $\gamma\ge2$. We shall see that the smoother the contact, the larger $\gamma$. Hence $\gamma$ is the parameter measuring what we called the softness. For a given strength and uniform mass density, it is easy to see that $\Delta v$ is inversely proportional to the length $L$ of the bar, while $\tau_\ell\propto L$. Hence, eq.\eqref{Ab} shows that the sound intensity increases as $L^{\gamma-2}$, on equal other conditions. 

Many studies deal with sound wave propagation in suitably \emph{constrained} bodies. In the realm of musical acoustics, for instance, the interest is focused on the wave characteristics, resulting from an initial strained configuration \cite{Benade}. Long-time contact forces on constrained bodies are considered, instead, in geophysical studies, as models for hearthquake generation \cite{BBC}. The dynamics of \emph{free} bodies, under the action of short-time contact forces, is usually approached as impact \cite{Strong, LNS} or bounching problems \cite{Cross, CBG}, with special reference to the resulting deformations. In focusing attention on \emph{long-time} contact forces in \emph{free} bodies, we deal with a problem that seems to have attracted less interest, apart from aereo-spatial engineering \cite{BIP, BA}. However, the relationship between beating and pushing just entails that kind of long-time dynamics. In this sense, our aim is primarly pedagogical and addresses to a readership of physics teachers, who wish to enrich the traditional studies on wave propagation in elastic structures, with non trivial elements on wave \emph{generation}.  

\section{The Thin Bar}
\label{Bar}

Figure 1 describes a thin elastic bar of mass $M$, transversal section area $a^2$, and unstressed length $L>>a$, free to move on a straight line. Neglecting the transversal strains, let $x\in[0,\:L]$ be the unstressed position of a section (relative to the bar's left edge) and $x'(x,\:t)$  be the actual position of the same section (relative to the laboratory frame of reference), so that 

\begin{equation}
\label{def.u}
u(x,\:t)=x'(x,\:t)-x
\end{equation}
\\
is the \emph{total} longitudinal displacement, including the part relative to the laboratory frame of reference. In the elastic approximation, the equation for $u(x,\:t)$ is:

\begin{subequations}
\label{eq.u,b.c.u}
\begin{equation}
\label{eq.u}
\partial^2_tu(x,\:t)-c_l^2\partial^2_x u(x,\:t)=\mathcal{F}(x,\:t)\:
\end{equation}
\\
($\partial_{\cdot\cdot\cdot}=$\:partial derivative), with border conditions:

\begin{equation}
\label{b.c.u}
\partial_x u(0,\:t)=\partial_x u(L,\:t)=0\:,
\end{equation}
\end{subequations}
\\
where $c_l$ is the longitudinal sound velocity and $\mathcal{F}(x,\:t)$ is a distribution of longitudinal forces per unit mass, directed along the bar. The border conditions eq.~\eqref{b.c.u} result from the vanishing of the strain $\partial_x u$ at the bar's ends, due to the absence of constraints \cite{LNS}. From eq.~\eqref{def.u} the centre of mass (c.m.) coordinate of the bar reads:

\begin{subequations}
\label{X(t)}
\begin{equation}
\label{X}
X(t)=\frac{\mu}{M}\int_0^La^2\mathrm{d}x\:u(x,\:t)+\frac{L}{2}\:,
\end{equation}
\\    
$\mu=M/(a^2L)$ being the (uniform) mass density. On time-differentiating the preceding equation twice, equations \eqref{eq.u,b.c.u} yield:

\begin{equation}
\label{eq.X}
\ddot{X}=\frac{\mu}{M}\int_0^La^2\mathrm{d}x\:\mathcal{F}(x,\:t)=\frac{1}{L}\int_0^L\mathrm{d}x\:\mathcal{F}(x,\:t)\:,
\end{equation}
\end{subequations}
\\
which recovers the obvious result that the c.m. moves under the action of the external forces' resultant (notice the importance of eq.~\eqref{b.c.u} in getting eq.~\eqref{eq.X}).

To satisfy the border conditions \eqref{b.c.u}, we write the solution of eq.~\eqref{eq.u} as:

\begin{equation}
\label{u}
u(x,\:t)=\sum_{n=-\infty}^\infty \rho_n(t)\cos\left(k_nx\right)\quad;\quad k_n=\frac{\pi n}{L}\:,
\end{equation}
\\
which yields\footnote{In a real bar, both $k_n$ and $\omega_n$ are upper limited by the Debye wave vector $k_D\propto1/d$ and Debye frequency $c_\ell k_D$, $d$ being a typical interatomic distance. Here we assume that those values are very large compared to all other quantities of interest.}:

\begin{equation}
\label{rhon}
\sum_{n=-\infty}^\infty \left[\ddot{\rho}_n(t)+\omega^2_n\rho_n(t)\right]\cos\left(k_nx\right)=\mathcal{F}(x,\:t)\:,\quad;\quad \omega_n=c_l\:k_n\:.
\end{equation}
\\
Let the contact force be concentrated at the left edge ($x=0$) of the bar, so that:

\begin{equation}
\label{Fstrano}
\mathcal{F}(x,\:t)=\delta(x)F(t)=\frac{F(t)}{2L}\sum_{n=-\infty}^\infty\cos(k_nx)\quad;\quad x\in[0,\:L]
\end{equation}
\\
($\delta(\cdot)=$ Dirac function\footnote{The second equality in \eqref{Fstrano} follows from the well known identity $\sum_{j=-\infty}^\infty\delta(x+2j\pi)=\sum_{n=-\infty}^\infty\cos(nx)/(2\pi)$.}). Equations \eqref{rhon} and \eqref{Fstrano} then yield:

\begin{equation}
\label{eq.rhok}
\ddot{\rho}_n(t)+\omega^2_n\rho_n(t)=\frac{F(t)}{2L}\:,
\end{equation}
\\
which reduces the problem to a set of independent harmonic oscillators, with the same time-dependent driving $F(t)/(2L)$ and different frequencies. The time-dependent driving is written in such a way that the \emph{duration}, \emph{strength} and \emph{shape} are clearly shown separately:

\begin{subequations}
\label{F,Psi}
\begin{equation}
\label{F}
\frac{F(t)}{2L}=
\begin{cases}
\frac{\Delta v}{2\tau}\Psi\left(\frac{t}{2\tau}\right)\:\:&\text{for }\:\:0\le t\le2\tau\\
&\\
0\:\:&\:\:\text{otherwise}
\end{cases}
\end{equation}
\\
with:

\begin{equation}
\label{intPsi}
\int_0^{1}\Psi(z)\mathrm{d}z =1\:.
\end{equation}
\end{subequations}
\\
From eq.s \eqref{F} and \eqref{Psi}, it follows that the duration of the contact force is $2\tau$, and the shape is determined by a normalized function $\Psi(\cdot)$. In what follows, we shall choose $\Psi(t/(2\tau))$ symmetric with respect to the mid-duration $\tau$ (see Fig. 2). Finally, we include the strength in the velocity achieved by the c.m. at the end of the contact, according to eq. \eqref{intPsi}:

\begin{equation}
\label{Deltav}
\Delta v:=\frac{1}{2L}\int_0^{2\tau}F(t)\mathrm{d}t=\dot{X}(2\tau)\:, 
\end{equation}
\\
where the second equality follows from the integration of eq.~\eqref{eq.X}, with the aid of eq.~\eqref{intPsi}. Choosing $\Delta v$ as the overall strength actually means choosing the \emph{total momentum} transferred by the contact force to the bar. This is convenient for many respects, provided one reminds that $\Delta v \propto L^{-1}\propto M^{-1}$ (eq.~\eqref{eq.X}).

Figure 2 illustrates some possible realizations of eq.s \eqref{F,Psi}. In particular, Figure 2(a) describes a hard contact/separation, as a constant contact force, starting and ending discontinuously at $t=0$ and $t=2\tau$. Figures 2(b) and 2(c) describe current models, applied to perfectly elastic bodies, where the initial increasing ramp is due to the elastic deformation of the impact region, which returns symmetrically to the equilibrium configuration (decreasing ramp), after reaching the maximum at $t=\tau$. The theoretical origin of this model is due to Hertz \cite{Hertz, McLG}, next developed by Hunter \cite{Hunter}, who suggested the \textquoteleft half-sine\textquoteright$\:$form, proportional to $\sin(\pi z)$ (Fig. 2(b)). A modified form, proportional to $[\sin(\pi z)]^{3/2}$, was introduced by Reed \cite{Reed} (Fig. 2(c)). Figures 2(b,c) usually refer to short-time impacts between \emph{hard} bodies (typically metallic). Extension to long times is just a choice of simplicity. As a term of comparison, Figure 2(d) describes a softer contact, with $\Psi(z)\propto \sin^2(\pi z)$ \cite{Tesi}.

\section{The single oscillator}
\label{1osc} 

Since equation \eqref{eq.rhok} reduces the problem of the internal motion of the bar to a set of independent harmonic oscillators, the present section will be concerned with the dynamics of a \emph{single} harmonic oscillator, with co-ordinate $\rho(t\:;\:\omega,\:\theta)$ and frequency $\omega$, under the action of the transient driving eq.~\eqref{F}:

\begin{subequations}
\label{rhosingle,theta}
\begin{align}
\ddot{\rho}(t\:;\:\omega,\:\theta)+&\omega^2\rho(t\:;\:\omega,\:\theta)=\frac{F(t)}{2L}\label{eq.rhosingle}\\
\nonumber\\
\theta&=\omega\tau\label{theta}
\end{align}
\end{subequations}
\\
In the time interval $[0,\:2\tau]$, we look for a solution of eq.~\eqref{eq.rhosingle} with initial conditions:

\begin{equation}
\label{in.cond.}
\rho(0\:;\:\omega,\:\theta)=0\quad,\quad\dot{\rho}(0\:;\:\omega,\:\theta)=0\:.
\end{equation}
\\
which follow from assuming the bar at rest and unconstrained for $t<0$. For $t>2\tau$, the free solution reads:

\begin{subequations}
\label{rho/inf,cont.cond.}
\begin{equation}
\label{rho/inf}
\rho_\infty(t\:;\:\omega,\:\theta)=\rho(2\tau\:;\:\omega,\:\theta)\mathrm{cos}(\omega t-2\theta)+\frac{\dot{\rho}(2\tau\:;\:\omega,\:\theta)}{\omega}\mathrm{sin}(\omega t-2\theta)\:,
\end{equation}
\\
with continuity conditions:

\begin{equation}
\label{cont.cond.}
\rho_\infty(2\tau\:;\:\omega,\:\theta)=\rho(2\tau\:;\:\omega,\:\theta)\quad,\quad\dot{\rho}_\infty(2\tau\:;\:\omega,\:\theta)=\dot{\rho}(2\tau\:;\:\omega,\:\theta)\:.
\end{equation}
\end{subequations}
\\
The quantities $\rho(2\tau\:;\:\omega,\:\theta)$ and $\dot{\rho}(2\tau\:;\:\omega,\:\theta)$ determine the elastic energy accumulated during the contact, and the way it transfers itself to the free harmonic oscillations at $t>2\tau$. If $\Psi(x)$ is expandible in a Fourier series, equation \eqref{rhosingle,theta} reads:

\begin{equation*}
\ddot{\rho}+\omega^2\rho=\sum_{j=0}^\infty\left[C_j\cos\omega_jt+S_j\sin\omega_jt\right]\:.
\end{equation*}
\\
The solution can be expanded in turn in a series of terms, each resulting from the well known resonant oscillator equation (without damping). The general solution of the \emph{homogeneous} harmonic equation ($F(t)=0$) must be added in order to satisfy the initial conditions \eqref{in.cond.}:

 \begin{equation*}
\rho(t)=\sum_{j=0}^\infty\left[V_j\cos\omega_jt+W_j\sin\omega_jt\right]+A\sin(\omega t+\phi)\:.
\end{equation*}
\\
It is left to the reader veryfing that the cases (a), (b) and (d), in Fig.2 yield\footnote {In case (d), remind that $\sin^2 \varphi/2=(1-\cos \varphi)/2$.}:

\begin{align*}
\rho(2\tau)&=\Psi_0\frac{\Delta v}{2\omega\theta}\left[1-\cos2\theta\right]\quad \text{(a)}\nonumber\\
\nonumber\\
&=\Psi_1\frac{\Delta v\pi}{2\omega\theta^2\left[1-(\pi/2\theta)^2\right]}\quad\text{(b)}\nonumber\\
\nonumber\\
&=\Psi_2\frac{\Delta v\pi^2}{4\omega\theta^3}\left[\cos 2\theta-\frac{1}{1-(\pi/\theta)^2}\right]\quad\text{(d)}\:.
\end{align*}
\\
Notice that in all cases $\rho(2\tau)$ is damped by a factor $\theta^{-(\alpha_0+1)}$, for $\theta>>1$, where $\alpha_0=0,\:1,\:2$ is the \emph{minimum exponent} of the expansion of $F(t)$ in powers of $t$. This crucial result can be extended to any $\Psi(z)$, expandible in a generic series of non-negative powers of $z$:

\begin{equation}
\label{Psi}
\Psi(z)=\sum_{j=0}^\infty\Psi_{\alpha_j}z^{\alpha_j}\quad;\quad 0\le\alpha_0<\alpha_1<\:\cdots\:,
\end{equation}
\\
provided the Fourier expansion is abandoned, in favor of the $t$-power expansion:

\begin{subequations}
\label{rho2,drho}
\begin{align}
\rho(t\:;\:\omega,\:\theta)&=-\frac{\Delta v}{2\omega\theta}\sum_{j=0}^\infty\Psi_{\alpha_j}\left(\frac{t}{2\tau}\right)^{\alpha_j}\sum_{n=1}^\infty\frac{\left[-(\omega t)^2\right]^n}{(\alpha_j+1)(\alpha_j+2)\cdots(\alpha_j+2n)}\label{rho2}\\
\nonumber\\
\dot{\rho}(t\:;\:\omega,\:\theta)&=-\frac{\Delta v}{2\omega t\theta}\sum_{j=0}^\infty\Psi_{\alpha_j}\left(\frac{t}{2\tau}\right)^{\alpha_j}\sum_{n=1}^\infty\frac{\left[-(\omega t)^2\right]^n}{(\alpha_j+1)(\alpha_j+2)\cdots(\alpha_j+2n-1)}\:,\label{drho}
\end{align}
\end{subequations}
\\
as obtained from eq.~\eqref{Psi} in the Appendix (eq. \eqref{App4}). So, for the sake of generality, and to account for the transient driving in Fig. 2(c) (a current model, in the impact theory), we shall use eq.s \eqref{rho2,drho} in what follows. The price to pay is a little journey into the realm of the Hypergeometric functions, which could look too technical for a non specialistic readership. To avoid a superflous effort, the calculations are thereby included in the Appendix, where the interested reader can find all details and, in particular, the demonstration of the general rule $\rho_\infty(2\tau,\:;\:\omega,\:\theta)\propto\theta^{-(\alpha_0+1)}$, to leading order in $\theta$,  which is of crucial importance for the aims of the present work.

\section{Internal displacements of the thin bar after the transient contact force}
\label{Bar2}

The passage from the single oscillator to the elastic thin bar, after the transient contact force, follows from eq.~\eqref{u}:

\begin{subequations}
\label{u2,thetan}
\begin{equation}
\label{u2}
u(x,\:t)=\sum_{n=-\infty}^\infty\rho_\infty(t\:;\:\omega_n,\:\theta_n)\cos(k_nx)\:,
\end{equation}
\\
\begin{equation}
\label{omegan,thetan}
\omega\rightarrow\omega_n=\frac{\pi}{\tau_l}n\quad;\quad\theta\rightarrow\theta_n=\frac{\pi\tau}{\tau_l}n\:,
\end{equation}
\end{subequations}
\\
where $\tau_l=L/c_l$ is the time a longitudinal elastic wave takes to cross the bar (recall the second eq.~\eqref{u} and \eqref{theta}). In view of the next developments, it is important to isolate the term $n=0\Rightarrow\theta_n=\omega_n=0$, in the series \eqref{u2}:

\begin{equation*}
\rho_\infty(t\:;\:0,\:0)=\Delta v\sum_{j=0}^\infty\Psi_{\alpha_j}\left[\frac{2\tau}{(\alpha_j+1)(\alpha_j+2)}+\frac{t-2\tau}{\alpha_j+1}\right]\:, 
\end{equation*}  
\\
as it follows from eq.s \eqref{rho3'A,drho3'A}  and \eqref{rho/inf}. On recalling eq.s~\eqref{F} and \eqref{Psi}, it is easy to see that the preceding equation becomes:

\begin{equation}
\label{Deltav2}
\rho_\infty(t\:;\:0,\:0)=\overbrace{\Delta v(t-2\tau)+\underbrace{\int_0^{2\tau}\mathrm{d}t_1\int_0^{t_1}\mathrm{d}t_2\frac{F(t_2)}{2L}}_{X(2\tau)}}^{X(t)}\:,
\end{equation}
\\ 
showing that the $0$-th term in the series \eqref{u2} is nothing but the free movement of the bar's c.m. after the applied contact force (eq.~\eqref{Deltav})\footnote{It is left to the reader showing that eq.~\eqref{Deltav2} can be obtained from eq.~\eqref{eq.rhok}, or eq.~\eqref{eq.rhosingle}, in the limit $\omega\rightarrow0$, without recursion to the series expansions eq.~\eqref{rho2,drho}.}. As obvious, the internal movements correspond to $n\ne0$, i.e. to $\omega_n,\:k_n\ne0$. We are especially interested in the \emph{internal} displacement velocity

\begin{subequations}
\label{vint}
\begin{equation}
v_{in}(x,\:t):=\underbrace{\partial_t u(x,\:t)}_{v(x,\:t)}-\dot{X}(t)\:,\label{vint1}
\end{equation}
\\
since the sound intensity will be shown to depend on $v_{in}^2(0,\:t)$ and $v_{in}^2(L,\:t)$ (Section \ref{soundint}). So, for $t>2\tau$, we differentiate equation \eqref{u2} with respect to $t$ and subtract $\Delta v=\dot{X}(2\tau)$. Since $\dot{\rho}_\infty(t\:;\:\omega,\:\theta)=$ $\dot{\rho}_\infty(t\:;\:-\omega,\:-\theta)$, on applying the decomposition formulas $2\cos \phi\cos \varphi=\cos(\phi+\varphi)+\cos(\phi-\varphi)$, $2\sin \phi\cos \varphi=\sin(\phi+\varphi)+\sin(\phi-\varphi)$, one gets:

\begin{align}
&v_{in}(x,\:t)=\sum_{n=-\infty}^\infty\dot{\rho}_\infty(t,\:;\:\omega_n,\:\theta_n)\cos(k_nx)-\Delta v=\text{ (from eq. \eqref{Deltav2}) }=\nonumber\\
\nonumber\\
&=\sum_{n=-\infty}^\infty\dot{\rho}_\infty(t,\:;\:\omega_n,\:\theta_n)\cos(k_nx)-\dot{\rho}_\infty(t,\:;\:0,\:0)=\text{ (from eq. \eqref{rho/inf}) }=\nonumber\\
\nonumber\\
&=\sum_{n=1}^\infty \Big\{\dot{\rho}(2\tau;\:\omega_n,\:\theta_n)\left[\cos(k_n(c_\ell \Delta t+x))+\cos(k_n(c_\ell \Delta t-x))\right]-\nonumber\\
&-\omega_n\rho(2\tau;\:\omega_n,\:\theta_n)\left[\sin(k_n(c_\ell \Delta t+x))+\sin(k_n(c_\ell \Delta t-x))\right]\Big\}\:,\label{vint2}
\end{align}
\\
\begin{equation}
\Delta t = t -2\tau
\end{equation}
\end{subequations}
\\
Equations \eqref{vint} show that the velocity of the internal displacements, after the end of the contact force, results in a superposition of progressive and regressive waves, running through the bar with velocity $c_\ell$, and responsible for the sound produced by the bar. Note that there is no sharp value of the parameters at which $\rho(2\tau;\:\omega_n,\:\theta_n)$ and $\dot{\rho}(2\tau;\:\omega_n,\:\theta_n)$ should vanish. In principle, even a delicate (and long lasting) pushing would thereby produce a \textquoteleft sound\textquoteright. How much loud the sound, is the question we shall answer to in the next sections.

\section{An upper limit for the internal velocity}
\label{tau->inf}

For the present aims, the study of the internal displacements' velocity $v_{in}(x,\:t)$ can be limited to a contact of duration much longer than the sound crossing time $\tau_l$. This corresponds to the limit $\theta>>1$, which makes it possible to approximate the next calculations up to the leading terms in $\theta$. From eq.s \eqref{rho3'A,drho3'A}, one has:

\begin{subequations}
\label{rho3',drho3'}
\begin{align}
\rho(2\tau\:;\:\omega,\:\theta)=&-\frac{\Delta v}{\theta^{\alpha_0+1}\omega}\mathcal{K}(\alpha_0)\:\cos\left(2\theta-\pi\alpha_0/2\right)\:+\mathrm{o}(\theta^{-\nu})\label{rho3'}\\
\nonumber\\
\dot{\rho}(2\tau\:;\:\omega,\:\theta)=&\frac{\Delta v}{\theta^{\alpha_0+1}}\mathcal{K}(\alpha_0)\:\sin\left(2\theta-\pi\alpha_0/2\right)\:+\mathrm{o}(\theta^{-\nu})\label{drho3'}\:,
\end{align}
\\
with

\begin{equation}
\label{nu}
\nu=\mathrm{min}\left\{\alpha_0+2\:,\:\alpha_1+1\right\}\:,
\end{equation}
\\

\begin{equation}
\label{strangeK}
\mathcal{K}(\alpha_0)=\frac{\Gamma\left(\frac{3+\alpha_0}{2}\right)\Gamma\left(\frac{4+\alpha_0}{2}\right)}{\sqrt{\pi}(\alpha_0+1)(\alpha_0+2)}\Psi_{\alpha_0}\:,
\end{equation}
\end{subequations}
\\ 
where $\Gamma(\cdot)$ is Euler's Gamma function\footnote{Euler's Gamma Function extends the factorial iteration rule $n!=n(n-1)!$ to a functional relationship $\Gamma(z+1)=z\Gamma(z)$, in all the complex plane. Tabulated values and formulas for $\Gamma(z)$ can be found, for instance, in ref.~\cite{Table}. A wide literature can be found on the web too.}. As anticipated in Section \ref{1osc}, equations \eqref{rho3',drho3'} show that the initial conditions, determining the free oscillation amplitudes at $t>2\tau$, decrease as $1/\theta^{\alpha_0+1}$. It is important to notice that the larger the exponent $\alpha_0$, the smoother the raising of the contact force (eq.~\eqref{Psi}). So, $\alpha_0$ determines the softness of the initial contact, between the impacting agent and the oscillator.

From eq.s \eqref{u2}, \eqref{rho/inf} and \eqref{rho3',drho3'}, one gets the following inequalities:

\begin{align}
|v_{in}(x,\:t)|&\le2\sum_{n=1}^\infty\left|\dot{\rho}_\infty(t\:;\:\omega_n,\:\theta_n)\right|<\nonumber\\
&<4\sum_{n=1}^\infty\left[\left|\rho(2\tau\:;\:\omega_n,\:\theta_n)\right|+\frac{\left|\dot{\rho}(2\tau\:;\:\omega_n,\:\theta_n)\right|}{\omega_n}\right]\omega_n<\nonumber\\
&<8\Delta v\left|\mathcal{K}(\alpha_0)\right|\sum_{n=1}^\infty\left[\frac{1}{(\theta_n)^{\alpha_0+1}}+\mathrm{o}(1/\theta_n^\nu)\right]\:.\nonumber
\end{align}
\\
Neglecting higher order terms in $\theta_n$, the preceding inequalities yield:

\begin{equation}
|v_{in}(x,\:t)|<\overbrace{8\Delta v\left(\frac{\tau_\ell}{\pi\tau}\right)^{\alpha_0+1}\left|\mathcal{K}(\alpha_0)\right|\underbrace{\sum_{n=1}^\infty\frac{1}{n^{\alpha_0+1}}}_{\zeta(\alpha_0+1)}}^{v_M(\alpha_0,\:\Delta v,\:\tau)}\:,\label{vmax}
\end{equation}
\\
where $\zeta(\cdot)$ is Riemann's Zeta function\footnote{Riemann's Zeta Function represents an amazing mathematical object, when analytically extended to all the complex plane. Among other subtleties, it contains Riemann's Hypothesis, one of the still unsolved problems in Mathematics.} and use has been made of the relationships \eqref{omegan,thetan}. If $\alpha_0=0$, equation \eqref{vmax} is useless, since $\lim_{x\rightarrow1}\zeta(x)=\infty$. However, it is possible to deal with the case $\alpha_0=0$ in a different way. Since $\mathcal{K}(0)=1/4$, from eq.s~\eqref{rho3',drho3'} and \eqref{rho/inf}, one gets:

\begin{equation*}
\rho_\infty(t,\:;\:\omega_n,\:\theta_n)=-\frac{\Delta v\Psi_0}{4\omega_n\theta_n}\cos(\omega_nt)+\mathrm{o}\left(\theta_n^{-\nu}\right)\quad(\alpha_0=0)\:
\end{equation*}
\\
and from eq.s~\eqref{vint}:

\begin{align}
&v_{in}(x,\:t)=\frac{\Delta v\Psi_0}{4}\sum_{n=1}^\infty\left\{\frac{\left[\sin(k_n(c_\ell t+x))+\sin(k_n(c_\ell t-x))\right]}{\theta_n}+\mathrm{o}\left(\theta_n^{-\nu}\right)\right\}=\nonumber\\
\nonumber\\
&=\frac{\Delta v\Psi_0}{4}\left(\frac{\tau_\ell}{\pi\tau}\right)\sum_{n=1}^\infty\Big\{\frac{1}{n}\left[\sin\left(\frac{\pi}{L}n(c_\ell t+x)\right)+\sin\left(\frac{\pi}{L}n(c_\ell t-x)\right)\right]+\label{vint0}\\
&+\mathrm{o}\left(n^{-\nu}\right)\Big\}\quad(\alpha_0=0)\:.\nonumber
\end{align}
\\
The two series in r.h.s. of eq. \eqref{vint0} correspond to the Fourier expansions of two periodic toothsaw functions, whose values range between $-\pi/2$ and $3\pi/2$ \cite{Wolf2}. So, neglecting higher order terms, equation \eqref{vint0} yields:

\begin{equation}
\label{vmax0}
\left|v_{in}(x,\:t)\right|<\frac{3\Delta v\Psi_0}{4}\left(\frac{\tau_\ell}{\tau}\right)=v_M(0,\:\Delta v,\:\tau)\quad(\alpha_0=0)\:.
\end{equation}
\\

In conclusion, equations \eqref{vmax} and \eqref{vmax0} show that the upper limiting value for the velocity of the internal displacements reads, in general:

\begin{subequations}
\label{vmax,R}
\begin{equation}
\label{vmaxgen}
v_M(\alpha_0,\:\Delta v,\:\tau)=\Delta v\left(\frac{\tau_\ell}{\tau}\right)^{\alpha_0+1}\mathcal{R}(\alpha_0)
\end{equation}
\\
where:

\begin{equation}
\label{R}
\mathcal{R}(\alpha_0)=
\begin{cases}
\frac{3}{4}\Psi_0&\:;\quad\alpha_0=0\\
\\
8\left|\mathcal{K}(\alpha_0)\right|\zeta(\alpha_0+1)&\:;\quad\alpha>0\:.
\end{cases}
\end{equation}
\end{subequations}
\\

\section{The sound intensity}
\label{soundint}
In the preceding sections, the bar was assumed to move in the vacuum. Here we introduce the atmospheric gas, in which the sound propagates, in order to study the sound intensity $\mathcal{I}_{snd}$, i.e. the power per unit area transferred from the bar to the gas, after the contact force eq.~\eqref{Fstrano}. The data referring to the bar's structure, the atmospheric gas and the contact force are reported in Table 1.  

\begin{table}[H]
\centering
\begin{tabular}{|c|c|c|c|c|}
\multicolumn{5}{c}{}\\
\multicolumn{5}{c}{\textbf{Steel Bar Data}} \\
\hline
&&&&\\
$L$ $(\text{\footnotesize{m}})$ & $a$ $\text{(\footnotesize{m}})$ & $M^*$ $(\text{\footnotesize{Kg}})$ & $c_\ell^*$ $(\text{\footnotesize{m/s}})$ & $\tau_\ell$ $(\text{\footnotesize{s}})$ \\
$\text{\footnotesize{length}}$ & $\text{\footnotesize{thickness}}$ & $\text{\footnotesize{mass}}$ & $\text{\footnotesize{sound velocity}}$ & $\text{\footnotesize{sound crossing time}}$\\
&&&&\\
\hline
&&&&\\
0.2 & 0.01 & 0.16 & $5.9\times10^3$ & $3.33\times10^{-4}$ \\
&&&&\\
\hline
\multicolumn{5}{c}{}\\
\multicolumn{5}{c}{\textbf{Room Atmosphere Data}} \\
\hline
&&&&\\
$<m>^*$ $\text{(\footnotesize{Kg}})$ & $T$  $(^{\text{\footnotesize{o}}}\text{\footnotesize{K}})$ & $P_{at}$ \footnotesize{(Pa)} & $\eta^*$ $(\text{\footnotesize{N s/m}}^{\text{\footnotesize{2}}})$ & $v_T$ $(\text{\footnotesize{m/s}})$ \\
\footnotesize{mean molecular mass} & \footnotesize{temperature} & \footnotesize{pressure} & \footnotesize{viscosity} & \footnotesize{thermal velocity} \\
&&&&\\
\hline
&&&&\\
$5\times10^{-26}$ & 300 & $10^{5}$ & $10^{-5}$ & $1.87\times10^2$ \\
&&&&\\
\hline
\multicolumn{3}{c}{}\\
\multicolumn{3}{c}{\textbf{Contact Force Data}} \\ 
\cline{1-3}
&&\\  
 $<f>$ $(\text{\footnotesize{N}})$ & $\tau$ $(\text{\footnotesize{s}})$ & $\Delta v$ $(\text{\footnotesize{m/s}})$ \\
\footnotesize{mean strength applied} & \footnotesize{contact force half duration} & \footnotesize{c.m. velocity} \\
&&\\
\cline{1-3}
&&\\
1 & $8\times10^{-2}$ & 1 \\
 &&\\
 \cline{1-3}
\end{tabular}
\caption{Entries relative to the bar, the atmospheric gas and the contact force. $(^*)$ Data from ref. \cite{HCP}.} 
\end{table} 

All internal dissipation processes are neglected, as is the case for an ideal elastic medium. The transversal strains are neglected too, so we consider the sound produced by the longitudinal motion of the bar's edges $u(0,\:t)$ and $u(L,\:t)$ (of area $a^2$), moving with velocities $v(0,\:t)$ and $v(L,\:t)$ (recall eq.~\eqref{vint1}), with respect to the atmospheric gas, in room conditions. The edge's velocity eq.~\eqref{vint1} results from a slow component $\dot{X}(t)$ (the velocity of the c.m. (eq.~\eqref{Deltav2})), plus a high frequency component $v_{in}(\lambda,\:t)$ ($\lambda=0,\:L$), which is expected to average out to zero, during the time scale of change of $\dot{X}(t)$. Hence the effects of the atmospheric gas on the dynamics of c.m. and internal displacements can be treated separately.  In particular, the c.m. velocity is damped by the gas viscosity $\eta$, while the rapid oscillations of the edges transfer the bar's internal energy to the gas, as pressure waves, i.e. as \emph{sound}. In normal conditions, $|\Delta v|$ and $|v_{in}(\lambda,\:t)|$ are very small, compared to the average velocity of the gas particles. This makes it possible to assume a laminar regime for the motion of the bar in the atmosphere, and to apply the results of ref. \cite{BC, G} for the pressure waves. First, let us calculate the damping of the c.m. velocity $\dot{X}(t)$, on neglecting the rapid oscillations of the front edge $u(L,\:t)$. According to Stoke's equation, one has, after the contact force ($t>2\tau$):

\begin{equation}
\ddot{X}=-\underbrace{R\frac{\eta a}{M}}_{\varpi}\dot{X}\:,
\end{equation}
\\
which defines the damping rate $\varpi$ of the c.m. velocity, in terms of the mass $M$, the linear size $a$ of the front edge and the viscosity $\eta$ (Table 1). The numerical factor $R\sim1\div10$ depends on the shape of the front edge. It is left to the reader verifying the condition $\varpi\tau_\ell<<1$, which ensures that the rapid oscillations do actually average out to zero on the damping time scale.

As for the sound generation, we treat the edges as pistons, moving with velocity $v_{in}(\lambda,\:t)$ ($\lambda=0,\:L$), and apply equation (7) of ref. \cite{BC} for the pressure exerted by the gas:

\begin{align}
&P(\lambda)=P_{at}\left\{1-\frac{8}{\pi}\left[(-1)^{\lambda/L}\left(\frac{v_{in}(\lambda,\:t)}{v_T}\right)-\left(\frac{v_{in}(\lambda,\:t)}{v_T}\right)^2\right]\right\}\quad(\lambda=0,\:L)\:,\label{Plambda}
\end{align} 
\\
where $P_{at}$ is the equilibrium atmospheric pressure and $v_T=\sqrt{4\kappa T/(3\pi <m>)}$ is a thermal velocity, characterizing the air particles of average mass $<m>$ (Table 1). The factor $(-1)^{\lambda/L}$ accounts for the fact that the left edge ($\lambda=0$) and the right edge ($\lambda=L$) exert, respectively, an expansion or a compression, when their velocity is positive. The force on each point of the edges' surface, contrasting the edges' velocity, is \footnote{It is obvious that the very small effect of $P_{at}$, on the bar at rest and unforced, has been neglected.}:

\begin{subequations}
\label{flambda}
\begin{equation}
f(\lambda,\:t)=(-1)^{\lambda/L}a^2P_{at}\:.\label{flambda}
\end{equation}
\\
Hence, the motion equation of the internal displacements $u_{int}(x,\:t):=u(x,\:t)-X(t)$, for $t>2\tau$, becomes:

\label{eq.mot.sound}
\begin{equation}
\label{eq.mot.sound1}
\partial_t^2u_{int}(x,\:t)-c_\ell^2\partial_x^2u_{int}(x,\:t)=\frac{f(x,\:t)}{\mu}\left[\delta(x)+\delta(x-L)\right]
\end{equation}
\\
or, equivalently, on account of eq. \eqref{vint1}:

\begin{align}
&\overbrace{\frac{\mathrm{d}}{\mathrm{d}t}\int_0^La^2\mu\:\mathrm{d}x\:\frac{(v_{in}(x,\:t))^2+c_\ell^2(\partial_xu_{int}(x,\:t))^2}{2}}^{\dot{E}_{in}}=\nonumber\\
\nonumber\\
&=\int_0^L\mathrm{d}x\:f(x,\:t)v_{in}(x,\:t)\left[\delta(x)+\delta(x-L)\right]=\label{eq.mot.sound}\\
\nonumber\\
&=f(0,\:t)v_{in}(0,\:t)+f(L,\:t)v_{in}(L,\:t)
\:,\nonumber
\end{align}
\end{subequations}
\\
where $\mu$ is the (uniform) mass density and $E_{in}$ is the internal energy of the bar. Equation \eqref{eq.mot.sound} shows that the \emph{internal} power per unit area $-\dot{E}_{in}/a^2$, transferred by the bar to the atmospheric gas, is:

\begin{subequations}
\label{I,IM}
\begin{align}
\label{I}
&\mathcal{I}_{snd}(t)=-\frac{\left[f(0,\:t)v_{in}(0,\:t)+f(L,\:t)v_{in}(L,\:t)\right]}{a^2}=\text{ (from eq.s \eqref{flambda}) }=\nonumber\\
\nonumber\\
&=P_{at}\frac{8}{v_T\pi}\left[\left(v_{in}(0,\:t)\right)^2+\left(v_{in}(L,\:t)\right)^2+\mathrm{o}\left(v_{in}^3/v_T\right)\right]\:.
\end{align}
\\
Since the internal velocity fades out in time, due to the energy loss described by eq.~\eqref{eq.mot.sound}, the value $v_M$ (eq.~\eqref{vmaxgen}), calculated in the vacuum, can be used \emph{a fortiori} as an upper limit for the internal velocity in the atmospheric gas. Hence, neglecting higher-order terms, equation \eqref{I} yields:

\begin{equation}
\label{IM}
\mathcal{I}_{snd}(t)<\mathcal{I}_M\left(\alpha_0\right)=\underbrace{P_{at}\frac{16}{v_T\pi}(\Delta v)^2\left(\frac{\tau_\ell}{\tau}\right)^{2(\alpha_0+1)}}_{\mathcal{I}_0}\mathcal{R}^2(\alpha_0)\:.
\end{equation}
\end{subequations}
\\
Inserting the data from Table 1 (in particular $\tau_\ell/\tau=0.417\times10^{-3}$) and the numerical factors $\mathcal{R}(\alpha_0)$ (Table 2) in eq.~\eqref{IM}, one finally gets the results reported in Table 2 for $\mathcal{I}_M\left(\alpha_0\right)$. 

\begin{table}[H]
\centering
\begin{tabular}{|c|c|c|c|c|c|c|c|}
\hline
\hline
\
&&&&&&\\
& $\alpha_0$ & $\Psi_{\alpha_0}$ & $\mathcal{R}$ & $\mathcal{I}_M$ $(\text{\footnotesize{W/m}}^2)$ & $\mathcal{I}_M$ $(\text{\footnotesize{dB}})$ & \textbf{Human ears'} \\
&&&&&& \textbf{perception}\\
&&&&&&\\
\hline
\hline
&&&&&&\\
1-a & 0 & 1 & $3/4$ & $1.66\times10^{-5}$ & 72 & limit of disturbance \\
&&&&&& \small{(1 meter far)}\\
&&&&&&\\
\hline
&&&&&&\\
1-b & 1 & $\pi^2/2$ & $\pi^4/12$ & $3.38\times10^{-10}$ & 25 & \small{wispers} \\
&&&&&&\small{(1 meter far)} \\
&&&&&&\\
\hline
&&&&&&\\
1-c & 3/2 & $\frac{3\sqrt{2}\pi^3}{\Gamma^2(1/4)}$ & $\frac{9\pi^{7/2}\zeta(5/2)}{4\Gamma^2(1/4)}$ & $3.39\times10^{-13}$ & -5 & \footnotesize{inaudible}\\
&&&&&&\\
\hline
&&&&&&\\
1-d & 2 & $2\pi^2$ & $2\pi^2\zeta(3)$ & $5.0\times10^{-16}$ & -33 & \scriptsize{inaudible} \\
&&&&&&\\
\hline
\end{tabular}
\caption{Intensity $\mathcal{I}_M(\alpha_0)$ resulting from the transient contact forces in Fig. 2. The third and fourth columns report the analytical expressions of the entries (eq.s~\eqref{IM}, \eqref{R},\eqref{Psi}). Seventh column from \cite{Wiki}.} 
\end{table}

\section{Discussion and conclusions}
\label{DC}

The main task of the present work is fully pedagogical and results in providing a formal representation of what one currently calls \textquoteleft beat\textquoteright$\:$and \textquoteleft push\textquoteright. A perfectly elastic, thin metallic bar is chosen as model system, for simplicity reasons. The elements that determine the sound/movement response are the \emph{strength}, the \emph{duration} and the \emph{softness} of the transient contact force $\mathcal{F}(x,\:t)$, applied to one edge of the bar ($x=0$). The strength can be conveniently included in the velocity $\Delta v$ attained by the c.m. at the end of the contact force. In fact, the sound intensity $\mathcal{I}_{snd}$ is proportional to $\Delta v^2$, as shown by eq.~\eqref{IM}. The duration $2\tau$ enters through a coefficient $(\tau_\ell/\tau)^\gamma$, where $\tau_\ell$ is the time the (longitudinal) sound takes to cross the bar. What we called the \textquoteleft softness\textquoteright$\:$ is represented by the positive exponent $\gamma=2(\alpha_0+1)$. Actually, the non negative number $\alpha_0$ is the smallest exponent in a series expansion of the contact force $\mathcal{F}(x,\:t)=\sum_{n=0}^\infty \mathcal{F}_n(x)t^{\alpha_n}$ and, thereby, determines the smoothness of the initial contact between the external agent and the bar. 

It is found that there is no sharp separation between beat (sound production) and push (overall movement): the distinction one is used to do, between those everyday actions, is just a matter of body perception. Actually, it is a common experience that beating a suspended object does not simply result in a clang, but also makes the c.m. move. What might look less obvious is that even a delicate and long lasting push does produce a \textquoteleft sound\textquoteright, so weak, however, to be inaudible to human ears, as shown in Table 2. The results reported therein refer to an average strength of $1\:\mathrm{N}$, making the bar's c.m. attain a velocity $\Delta v=1\:\mathrm{m/s}$, after $2\tau=1.6\times10^{-1}\mathrm{s}$ (Table 1). The sound intensity (at the source) ranges between a noisy level ($\alpha_0=0$) to a quite inaudible level ($\alpha_0=2$), depending on the shape in time of the contact force (Fig. 2), which determines the softness parameter $\alpha_0$.

An important point to be stressed is that, for $\tau_\ell/\tau<<1$, the factor $\mathcal{I}_0$ in the inequality eq.~\eqref{IM} is the \emph{exact} coefficient of $\mathcal{I}_{snd}(t)$, which multiplies an oscillating numerical function of time, whose amplitude is bounded from above by $\mathcal{R}^2(\alpha_0)$. Hence, one can use $\mathcal{I}_0$ as an exact expression, when referring to the dependence of $\mathcal{I}_{snd}(t)$ on the physical parameters reported in Table 1. In particular, since $\tau_\ell\propto L$ and $\Delta v\propto L^{-1}$, equation \eqref{IM} shows that the sound intensity increases with the length as $L^{2\alpha_0}$. So, a longer object sounds louder\footnote{Apart from the critical case $\alpha_0=0$ (Fig. 1-a), corresponding to a hard contact. In this case the sound intensity turns out to be independent of the length $L$.} and moves slower than a shorter one, the material, the transversal section and the applied contact force being the same. This effect depends essentially on the softness parameter $\alpha_0$ and, thereby, could be used to determine experimentally the incipient shape (in time) of the contact force.

\begin{appendices}
\numberwithin{equation}{section}
\section{Appendix}

The present Appendix is devoted to the technical details about the solution of the differential problem :

\begin{equation}
\ddot{\rho}(t)+\omega^2\rho(t)=\frac{\Delta v}{(2\tau)^{\alpha_j+1}}\sum_{j=0}^\infty \Psi_{\alpha_j}t^{\alpha_j}\:. \label{App0}
\end{equation}
\\
resulting from eq.s (11a), (9a) and (16) of the text. Such solution can be written as follows:

\begin{subequations}
\label{App}
\begin{equation}
\rho(t)=\sum_\alpha\rho_\alpha(t)\:,\label{App1}
\end{equation}
\\
with:

\begin{equation}
\ddot{\rho}_\alpha+\omega^2\rho_\alpha=A_\alpha t^\alpha\:\label{App2}
\end{equation}
\\
\begin{equation}
A_\alpha:=\frac{\Delta v}{(2\tau)^{\alpha+1}}\Psi_\alpha\:\label{App3}
\end{equation}
\end{subequations}
\\
(index $j$ has been dropped from $\alpha_j$ for brevity). The solution of eq.~\eqref{App2}, with initial conditions $\rho_\alpha(0)=\dot{\rho}_\alpha(0)=0$, can be found by iteration, on setting:

\begin{align}
\rho_\alpha&=\underbrace{A_\alpha\frac{\alpha!}{(\alpha+2)!}t^{\alpha+2}}_{\rho_{\alpha,0}}+\rho_{\alpha,1}\Rightarrow\nonumber
\nonumber\\
&\Rightarrow\ddot{\rho}_{\alpha,1}+\omega^2\rho_{\alpha,2}=-A_\alpha\frac{\omega^2\alpha!}{(\alpha+2)!}t^{\alpha+2}\Rightarrow\nonumber\\
\nonumber\\
&\Rightarrow\rho_{\alpha,1}=-A_\alpha\frac{\omega^2\alpha!}{(\alpha+4)!}t^{\alpha+4}+\rho_{\alpha,2}\Rightarrow \cdots\nonumber\\
\nonumber\\
&\Rightarrow \rho_\alpha=-A_\alpha\sum_{n=1}^\infty\frac{(-1)^n\omega^{2(n-1)}\alpha!}{(\alpha+2n)!}t^{\alpha+2n}\:.\label{App4}
\end{align}
\\
Using eq.s \eqref{App4}, \eqref{App3}, \eqref{App1} and restoring the index $j$ in $\alpha$, it is easy to obtain:

\begin{equation}
\label{App3'}
\rho(t\:;\:\omega,\:\theta)=-\frac{\Delta v}{2\omega\theta}\sum_{j=0}^\infty\Psi_{\alpha_j}\left(\frac{t}{2\tau}\right)^{\alpha_j}\sum_{n=1}^\infty\frac{\left[-(\omega t)^2\right]^n}{(\alpha_j+1)(\alpha_j+2)\cdots(\alpha_j+2n)}\:,
\end{equation}
\\
i.e. the expression (17a) in the text. On recalling the definition of generalized Hypergeometric functions $_1F_2(a\:;\:b,\:c\:;\:z)$ (ref. [14]\footnote{For the passage from eq.~\eqref{App3'} to eq.s~\eqref{rho2A,drhoA}, use can be made of Mathematica (simply by typing the series as in eq.~\eqref{App3'}), in alternative to applying the definition of generalized Hypergeometric function in ref. [14].}), equation \eqref{App3'} yields:

\begin{subequations}
\label{rho2A,drhoA}
\begin{align}
\rho(t\:;\:\omega,\:\theta)&=\frac{\Delta v(\omega t)^2}{2\omega\theta}\sum_{j=0}^\infty\Psi_{\alpha_j}\left(\frac{t}{2\tau}\right)^{\alpha_j}\frac{_1F_2\left(1\:;\:\frac{3+\alpha_j}{2},\:\frac{4+\alpha_j}{2}\:;\:-\frac{(\omega t)^2}{4}\right)}{(\alpha_j+1)(\alpha_j+2)}\label{rho3A}\\
\nonumber\\
\dot{\rho}(t\:;\:\omega,\:\theta)&=\frac{\Delta v\:\omega t}{2\theta}\sum_{j=0}^\infty\Psi_{\alpha_j}\left(\frac{t}{2\tau}\right)^{\alpha_j}\frac{_1F_2\left(1\:;\:\frac{2+\alpha_j}{2},\:\frac{3+\alpha_j}{2}\:;\:-\frac{(\omega t)^2}{4}\right)}{(\alpha_j+1)}\:.\label{drhoA}
\end{align}
\end{subequations}
\\
On setting $t=2\tau$ in the preceding expressions, one has:

\begin{subequations}
\label{rho,drho/tau}
\begin{align}
\rho(2\tau\:;\:\omega,\:\theta)&=\frac{2\Delta v\:\theta}{\omega}\sum_{j=0}^\infty\Psi_{\alpha_j}\frac{_1F_2\left(1\:;\:\frac{3+\alpha_j}{2},\:\frac{4+\alpha_j}{2}\:;\:-\theta^2\right)}{(\alpha_j+1)(\alpha_j+2)}\label{rho/tau}\\
\nonumber\\
\dot{\rho}(2\tau\:;\:\omega,\:\theta)&=\Delta v\sum_{j=0}^\infty\Psi_{\alpha_j}\frac{_1F_2\left(1\:;\:\frac{2+\alpha_j}{2},\:\frac{3+\alpha_j}{2}\:;\:-\theta^2\right)}{(\alpha_j+1)}\:.\label{drho/tau}
\end{align}
\end{subequations}
\\
Recalling the basic asymptotic formula for the Hypergeometric functions \cite{Wolf} :

\begin{align}
_1F_2(1\:;\:b\:,\:c\:;\:z)&=\frac{\Gamma(b)\Gamma(c)}{2\sqrt{\pi}\Gamma(1)}(-z)^{(3/2-b-c)}\times\nonumber\\
\nonumber\\
&\times\left[\cos\left(\frac{\pi}{2}\left(\frac{3}{2}-b-c\right)+2\sqrt{-z}\right)+\mathrm{o}\left(\frac{1}{\sqrt{-z}}\right)\right]
\end{align} 
\\
($\Gamma(\cdot)$ = Euler Gamma function), one may extract from the sums in eq.s~\eqref{rho,drho/tau}, the leading terms, in the limit $\theta>>1$:

\begin{subequations}
\label{rho3'A,drho3'A}
\begin{align}
\rho(2\tau\:;\:\omega,\:\theta)=&-\frac{\Delta v}{\theta^{\alpha_0+1}\omega}\overbrace{\frac{\Gamma\left(\frac{3+\alpha_0}{2}\right)\Gamma\left(\frac{4+\alpha_0}{2}\right)}{\sqrt{\pi}(\alpha_0+1)(\alpha_0+2)}\Psi_{\alpha_0}}^{\mathcal{K}(\alpha_0)}\:\cos\left(2\theta-\pi\alpha_0/2\right)\:+\label{rho3'}\\
&+\mathrm{o}(\theta^{-\nu})\nonumber\\
\nonumber\\
\dot{\rho}(2\tau\:;\:\omega,\:\theta)=&\frac{\Delta v}{\theta^{\alpha_0+1}}\underbrace{\frac{\Gamma\left(\frac{2+\alpha_0}{2}\right)\Gamma\left(\frac{3+\alpha_0}{2}\right)}{2\sqrt{\pi}(\alpha_0+1)}\Psi_{\alpha_0}}_{\mathcal{K}(\alpha_0)}\:\sin\left(2\theta-\pi\alpha_0/2\right)+\label{drho3'}\\
&+\mathrm{o}(\theta^{-\nu})\nonumber
\end{align}
\\
\begin{equation}
\label{eta}
\nu=\mathrm{min}\left\{\alpha_0+2\:,\:\alpha_1+1\right\}\:.
\end{equation}
\end{subequations}
\\ 
Note that the identity between the overbraced coefficient in \eqref{rho3'} and the underbraced coefficient in \eqref{drho3'} follows from the functional relation $\Gamma(1+x)=x\Gamma(x)$. Equations \eqref{rho3'A,drho3'A} coincide with eq.s (21).

One might wonder why expanding $\Psi(x)$ in a series of (non negative) generalized powers $x^{\alpha_j}$, instead of a more familiar \emph{integer} powers expansion, that would look far easier. Actually, the equation:

\begin{equation}
\label{rhopol}
\Ddot{\rho}_j+\omega^2\rho_j=A_jt^j\:(j=0,\:1,\:\cdots)\:,
\end{equation}
\\
admits a j-degree \emph{polynomial} solution, instead of the infinite series eq.~\eqref{App4}. Seemingly, this procedure would skip the Hypergeometric functions, and could be applied even in the case $\Psi(x)\propto|\sin(\pi x)|^{3/2}$ (Fig. 2(c)), simply by expanding $\Psi(t/2\tau)$ about an instant different from zero (for instance, the mid-duration time $\tau$). However, such simplification is only apparent in any case, and turns into a relevant disadvantage when expanding about $t\ne0$. The reason is that one has to add a term $A\sin(\omega t+\phi)$ to the polynomial solution of eq.~\eqref{rhopol}, in order to satisfy the initial conditions (12) in MT. This means that $\rho_j(t)$ is actually represented by an \emph{infinite} series in $t$, which turns out to be just a special case of eq.s~\eqref{rho2A,drhoA}. Expanding about $t\ne0$, to avoid using non integer powers, entails a further problem. In this case, extracting the leading term $1/\theta^{\alpha_0+1}$, which refers to the expansion about $0$ and is the gist of the softness effect, becomes very complicated, if not impossible, without coming back to the Hypergeometric functions. 

As a final remark, we stress that the method used here to obtain the solution of the differential problem \eqref{App0} can be easily extended to all cases in which $\Psi(z)$ is \emph{piece-wise} expandible in a generalized power series. This could be useful in most concrete situations, in which the imperfect elasticity of the body makes the contact force asymmetric, with respect to the instant $\tau$ of maximum impact (see ref. [20]).
\end{appendices}
\\
\\
\emph{Aknowledgments}:
The author is grateful to the library of the Italian Air Force in Rome and, in particular, to CMS Gaetano Pasqua (PhD), for useful bibliographic suggestions. 
\\
\\

\end{document}